\documentclass[a4paper,fleqn]{cas-dc}

\usepackage[numbers, sort]{natbib}
\usepackage{cleveref}
\usepackage{graphicx}
\usepackage{subfig}
\usepackage{float}
\usepackage{tikz}
\usepackage{multirow}
\usepackage{placeins}
\usetikzlibrary{shapes.geometric, arrows}
\tikzstyle{startstop} = [rectangle, rounded corners, 
minimum width=3cm, 
minimum height=1cm,
text centered, 
draw=black, 
fill=red!30]

\tikzstyle{io} = [trapezium, 
trapezium stretches=true, 
trapezium left angle=70, 
trapezium right angle=110, 
minimum width=3cm, 
minimum height=1cm, text centered, 
draw=black, fill=blue!30]

\tikzstyle{process} = [rectangle,
minimum width=3cm, 
minimum height=1cm, 
text centered, 
text width=5cm, 
draw=black, 
fill=orange!30]

\tikzstyle{decision} = [diamond, 
minimum width=3cm, 
minimum height=1cm, 
text centered, 
draw=black, 
fill=green!30]
\tikzstyle{arrow} = [thick,->,>=stealth]

\def\tsc#1{\csdef{#1}{\textsc{\lowercase{#1}}\xspace}}
\tsc{DM}
\tsc{JPY}

\usepackage{lineno}

\begin{document}
\let\WriteBookmarks\relax
\def\floatpagepagefraction{1}
\def\textpagefraction{.001}

\shorttitle{Simulating the Angular Distribution of Cherenkov Radiation in Thick Media}

\shortauthors{D. Minchenko, J. P. Ya\~nez, A. Hallin}

\title [mode = title]{Simulating the Angular Distribution of Cherenkov Radiation in Thick Media}                      

\author[1]{Dmytro Minchenko}

\cormark[1]

\ead{minchenk@ualberta.ca}

\affiliation[1]{
organization={Dept. of Physics, University of Alberta},
city={Edmonton, Alberta},
postcode={T6G 2E1},
country={Canada}}

\author[1]{Juan Pablo Ya\~nez}

\author[1]{Aksel Hallin}

\cortext[cor1]{Corresponding author}

\begin{abstract}
We present a study of the emission of Cherenkov radiation in thick media, explore the limitations of the simulation tools currently in use in particle and nuclear physics, and propose a method for overcoming these limitations. We start with a derivation of the Cherenkov power spectrum and its angular profile, accounting for interference of the radiation emitted, in contrast with commonly used tools that assume perfect coherence. We then study the impact that the path of electrons through a medium has on the angular profile of Cherenkov light. Finally, we devise a model that can introduce these effects in Geant4 and tune it to explain calibration data from the water-phase of SNO+. We find that the tuned model significantly improves the agreement between data and simulation, so we provide it for its wider use.
\end{abstract}



\maketitle

\section{Introduction}

Cherenkov radiation~\cite{Cherenkov:1937mnd} is widely used to  detect and characterize charged particles, with applications in experimental particle physics, astrophysics, nuclear physics, and medicine. The Cherenkov light emitted by a particle depends upon particle type, energy and momentum. A property commonly used is that most light is emitted at a fixed angle $\theta_C$, also known as the Cherenkov angle.

The standard derivation of Cherenkov light emission \cite{Frank1991, Jackson} shows that a charged particle in a dielectric medium with refractive index $n$, moving on a straight line, with a constant velocity $\beta = \frac{v}{c}$ faster than the local speed of light in the medium, will emit Cherenkov light, a coherent front of light in a cone with half-angle $\theta$ given by
\begin{equation}
    \label{eq:cherenkov_angle}
    \cos{\theta_C} = \frac{1}{n \beta}.
\end{equation}
Since the Cherenkov angle is well-defined with respect to the direction of travel of the charged particle, the resulting cone of Cherenkov light can be used to trace the path of the particle with high precision. Experiments have exploited this effect for almost four decades to detect charged particles and infer their properties over a wide range of energies ~\cite{Seguinot:1993xz, Kolanoski:2020ksk}. The technique is so powerful that it will be continued to be used by future generations of detectors, which in turn requires ever more precise simulation of the process.

Modern simulation tools that study the movement of charged particles through matter often discretize physical processes. Interactions are simulated at scattering centers, with the particle undergoing free motion in-between them. The kinematics and probability of interactions are defined by scattering models that take into account a variety of physical phenomena.

Simulations of Cherenkov radiation are typically discretized. It is commonly assumed that the conditions required for emission of Cherenkov radiation persist along a charged particle's trajectory and the Cherenkov emission angle is calculated with respect to the average displacement of the particle. In reality, however, a particle is subject to a continuous interaction with the medium at the micron scale that deflects its trajectory and makes it lose energy continuously, alongside discrete interactions at atomic scales. As a result, the Cherenkov light emission might be significantly different from the predictions from simulations that approximate these effects. Cherenkov light is a coherent effect and we examine the potential interference and loss of coherence of the light front as well as the smearing of the emission angle that might result due to these effects.

This work was motivated by an observed tension between data and simulation reported by the SNO+ experiment when comparing the angular distribution of Cherenkov light emitted by electrons with MeV energies in water~\cite{SNO:2018ydj}. Internal studies explored the effects of multiple scattering \cite{Dedrick, Kobzev1979, kobzev1980spectral}, the coherence of the Cherenkov light emitted~\cite{BOWLER1996463}, and the influence of the simulation step size on the resulting Cherenkov light distribution \cite{BowlerEGS4}. However, none of these studies explain the tension. Therefore, in this work, we assess the approximations routinely made when simulating Cherenkov light, and propose a method to account for the effects that are currently neglected.

In Sec.~\ref{ch:derivation}, we provide a review of the classic derivation of Cherenkov light emission, including its associated approximations.  In Sec.~\ref{ch:scattering_models} we look at different methods for electron transport in simulation, and investigate their impact on their predicted angular distribution of Cherenkov light. In Sec.~\ref{ch:New_model}, we develop a new Cherenkov light radiation model, implement it in the commonly used Geant4 platform~\cite{AGOSTINELLI2003250}, and tune it to SNO+ calibration data. We find that, after tuning the model, the previously observed tension between the data and Monte-Carlo simulation is resolved.

\section{Cherenkov light from first principles}\label{ch:derivation}

We derive the radiated power of a charged particle as it scatters while traversing a medium\footnote{
We surveyed the literature searching for a derivation of the emission of Cherenkov light in thick media, but could not find any that would go beyond use of thin plates and straight-path approximations.}. This is based on the works of R.J.~Komar~\cite{Komar}, Schiff~\cite{Schiff}, and Dedrick~\cite{Dedrick}. Following Schiff, we define the current density $J_x$ of a point charge $e$ that moves with a constant speed $v$ along the z-axis starting at time $t=0$ and position $\textbf{r}=(x,y,z) = 0$:

\begin{equation}
    \label{Current_perp}
    J_x(\boldsymbol{r}, t) = J_y(\boldsymbol{r}, t) = 0,
\end{equation}
and
\begin{equation}
    \label{eq:Current_z}
    J_z(\boldsymbol{r}, t) = e\,v\,\delta(x)\cdot\delta(y)\cdot\delta(z - vt).
\end{equation}
where $\textbf{J}=(J_x,J_y,J_z)$ is the current density, $e$ the charge of the particle, and $v$ the velocity.  

To calculate the angular distribution of the Cherenkov radiation, we use the exact expression for the average energy radiated at a position $\boldsymbol{r}$ by a harmonically time-varying current distribution in a homogeneous isotropic dielectric medium \cite{Dedrick}, as
\begin{equation}
    \label{eq:Power}
    P_{k\omega}(\boldsymbol{r}) = \frac{n k^2}{2 \pi r^2 c} 
    \left| \int J_{\perp k} (\boldsymbol{r'}) \, e^{-in\boldsymbol{k} \cdot \boldsymbol{r'}}d\tau' \right|^2,
\end{equation}
where $P_{k\omega}$ is the energy flow per unit area and angular frequency in the direction of observation (parallel to $\boldsymbol{k}$ or $\boldsymbol{r}$), $|\boldsymbol{k}| = \omega / c$, $n$ is the index of refraction for the medium, $J_{\perp k}$ is the component of the current density perpendicular to $\boldsymbol{k}$ and $d\tau'$ is the volume element. Replacing the density in Eq.~\ref{eq:Current_z} by the Fourier amplitude of angular frequency $\omega$ we obtain
\begin{equation}
    \label{eq:Current_fourier}
    J_{z\omega}(\boldsymbol{r}, t) = \frac{e}{2 \pi} \, \delta(x)\cdot\delta(y)\,e^{(iwz/v)}.
\end{equation}
Substituting $J_{\perp k} = J_{z \omega} \sin{\theta}$ into Eq.~\ref{eq:Power} we obtain the energy flow per unit area and angular frequency
\begin{equation}
    \label{eq:Power_res}
    2 \pi P_{k\omega}(\boldsymbol{r}) = \frac{n\, e^2 \omega^2 \sin^2{\theta}}{4 \pi^2 r^2 c^3} 
    \left| \int e^{i\omega z'(\frac{1}{v} - \frac{n \cos{\theta}}{c})}dz' \right| ^2.
\end{equation}

For a path length $L$ centered at the origin, the integral is 
\begin{equation}
    \label{eq:int_res}
    \int^{L/2}_{-L/2} e^{i\omega z'\left(\frac{1}{v} - \frac{n\cos{\theta}}{c}\right)}dz' =
    \frac{ 2 \sin{ \left[ \frac{\omega L}{2}\left(\frac{1}{v} - \frac{n\cos{\theta}}{c}\right) \right] }} {\omega\left(\frac{1}{v} - \frac{n\cos{\theta}}{c}\right)}.
\end{equation}
For a particle with a straight and infinite trajectory we evaluate the limit for infinite $L$ to be  $2\pi \delta \left( \frac{1}{v} - \frac{n \cos{\theta}}{c} \right),$ leading to the classical expression shown in Eq.~\ref{eq:cherenkov_angle}.

From Eq.~\ref{eq:Power_res} the radiated power is proportional to
\begin{equation}
\label{eq:ang_shape1}
L^2 \sin^2{\theta} \frac{\sin^2{\chi}}{\chi^2}
\end{equation}
with
\begin{equation}
    \label{eq:ang_shape2}
\chi = \frac{\omega L}{2} \left( \frac{1}{v}- \frac{n \cos{\theta}}{c} \right) \equiv
\frac{\pi L}{\lambda'} \left( \frac{1}{n\beta}- \cos{\theta} \right),
\end{equation}
where $\lambda' = 2\pi c/n\omega$ is the wavelength in the medium. 

The behavior of Eq.~\ref{eq:ang_shape1} depends on how $L$ compares to $\lambda'$. In the case that $L \ll \lambda'$, the $\sin^2{(\chi)}/\chi^2$ in Eq.~\ref{eq:ang_shape1} becomes constant and the radiation is emitted over a dipole angular distribution. For $L \gg \lambda'$, the shape of Eq.~\ref{eq:ang_shape1} resembles a gaussian function summed to a cosine function with a relatively smaller amplitude, and it is valid in the range $\chi=[-\infty,+\infty]$. The total light output is given by the integral, and in this limiting case about 90\% of the contribution comes from the range $\chi=[-\pi,+\pi]$. Here, the angular distribution of the radiation emitted is sharply peaked at the Cherenkov angle $\theta_C$ and has a full width at half maximum
\begin{equation}
    \delta \theta \simeq \frac{\lambda'}{L\sin{\theta_C}}.
\end{equation}

We note that the integral of Eq.~\ref{eq:ang_shape1} can be done in $\chi$ space, where the valid range of $\chi$ depends linearly on $L/\lambda'$. By requiring that $|\chi|$ can acquire values larger than $pi$, motivated by the limiting case of  $L \gg \lambda'$, we can estimate the regime where $L$ is sufficiently larger than $\lambda'$ to achieve coherence. Since the minimum value of $\chi$ (where $\theta=0$) is always the most restrictive one, we use it and demand it is $\leq-\pi$, which results in the coherence condition that
\begin{equation}
    \label{eq:coh_cond}
    \frac{L}{\lambda'} \left(1- \frac{1}{n\beta} \right) \geq 1.
\end{equation}
Water is a commonly used medium for experiments involving Cherenkov radiation. Using a refractive index for water of $n = 4/3$ and assuming that $\beta = 1$, Eq.~\ref{eq:coh_cond} becomes
\begin{equation}
    L > 4 \lambda'.
\end{equation}
Cherenkov detectors are typically sensitive to light with wavelengths $300 $~nm $<\lambda'$ $ < 720$~nm, so full coherence requires straight path lengths between 1.2-3~$\mu$m. This is comparable to the mean free path between scattering of electrons in the water, approximately~$1.3~ \mu$m~\cite{EGS4}.

Dedrick \cite{Dedrick} derives the formula including interference between light emitted by different track segments. \Cref{eq:Power} becomes:

\begin{equation}
    \label{eq:power_sum}
    P_{k\omega}(\boldsymbol{r}) = \frac{n k^2}{2 \pi r^2 c }
    \left| \sum^N_{\nu = 1} I_{\nu} \right| ^2,
\end{equation}
where the contribution of each segment $\nu$ is

\begin{equation}
    \label{eq:I_nu}
    I_{\nu} = \frac{e}{2 \pi} \sin{\Theta_{\nu}} e^{i(\delta_{\nu} + \chi_{\nu})} 
    l_{\nu} \frac{ \sin{\chi_{\nu}} }{i\chi_{\nu}}
\end{equation}
with the phase angles given by
\begin{equation}
    \chi_{\nu} = \frac{\omega l_{\nu}}{2} \left( \frac{1}{v_{\nu}} - \frac{n}{c} \cos{\Theta_{\nu}} \right) \equiv \frac{\pi l_{\nu}}{\lambda'} \left( \frac{1}{n \beta_{\nu}} - \cos{\Theta_{\nu}} \right),
\end{equation}
with
\begin{equation}
    \cos{\Theta_{\nu}} = \cos{\theta} \cos{\theta_{\nu}} + \sin{\theta} \sin{\theta_{\nu}} \cos{(\varphi - \varphi_{\nu})}
\end{equation}
and
\begin{equation}
    \delta_{\nu} = \omega t_{\nu} - nk (x_{\nu}\sin{\theta}\cos{\varphi} + 
    y_{\nu} \sin{\theta}\sin{\varphi} + z_{\nu} \cos{\theta}).
\end{equation}
Here $l_{\nu}$ is the length of the $\nu$ step from $x_{\nu}$, $y_{\nu}$, $z_{\nu}$ to $x_{\nu + 1}$, $y_{\nu + 1}$, $z_{\nu + 1}$, $\theta_{\nu}$ and $\varphi_{\nu}$ are the polar angles of the segment, and the particle is at point $\nu$ at time $t_{\nu}$ with the speed $v_{\nu}$.

The expression for the total energy radiated is composed both of terms $I_{\nu} I^*_{\nu}$ and also terms
$(I_{\nu} I^*_{\mu} + I^*_{\nu} I_{\mu})$ that represent interference effects between segments.

Our step-based Monte Carlo simulation uses Eq.~\ref{eq:power_sum} to calculate the total angular distribution of Cherenkov light.

\section{Electron transport}
\label{ch:scattering_models}

The distribution of electron directions and step sizes has a significant impact on the angular distribution of Cherenkov light. We focus on the difference between single-scattering and multiple-scattering models.

\subsection{Charged particle scattering models}

Monte-Carlo scattering models define the longitudinal and lateral displacement of a particle along its trajectory, forming steps. Models can be divided into two groups, condensed and detailed~\cite{Urban}. Detailed algorithms, also known as ``single scattering models'', simulate each interaction and therefore provide the highest accuracy possible. Condensed models, on the other hand, average several physical interactions to calculate displacement and energy losses, typically reducing the number of steps by factors of hundreds, therefore reducing the computational needs by a similar factor. Condensed models are commonly referred to as ``multiple scattering models''. 

Commonly used multiple scattering models in Geant4 are Urban~\cite{Urban}, Penelope~\cite{Penelope} and Wentzel~\cite{Ivanchenko_2010}. While the step size of the condensed model can be shortened artificially to increase the precision of the simulation, the most accurate simulation
is obtained when using a detailed Single Scattering model~\cite{Ivanchenko_2010}. 

We simulate the Cherenkov light angular distribution for MeV electrons in water with Geant4. We use the Urban model~\cite{Urban} for the multiple scattering case, and the model described in~\cite{Ivanchenko_2010} for the single scattering case. We inject 1000 electrons in the middle of a 6~m radius water sphere with an initial momentum direction along the z-axis. The initial electron energies are 0.3~MeV, 0.5~MeV, 1~MeV, and 5~MeV. We take into account bremsstrahlung, ionization, scattering and the Cherenkov light emission process. 

The photons propagate in straight paths through the water, with all physical processes disabled, until they are absorbed at the sphere's surface. The position of the photons on the sphere with respect to its center is used to compute the total Cherenkov light angular distribution far from its emission point. 
From these distributions, we compute three useful quantities that allow us to compare the results from different simulations, namely the angle at which the maximum number of photons are emitted, the fraction of photons emitted in the tails, with angles $> \pi / 2$, and the Full Width Half Maximum (FWHM) of the resulting Cherenkov light distribution. 

\subsection{Multiple Scattering vs Single Scattering}
\label{ch:MSvsSS}

\begin{figure*}[!tbh]
  \centering
  \subfloat{\includegraphics[width=0.49\textwidth]{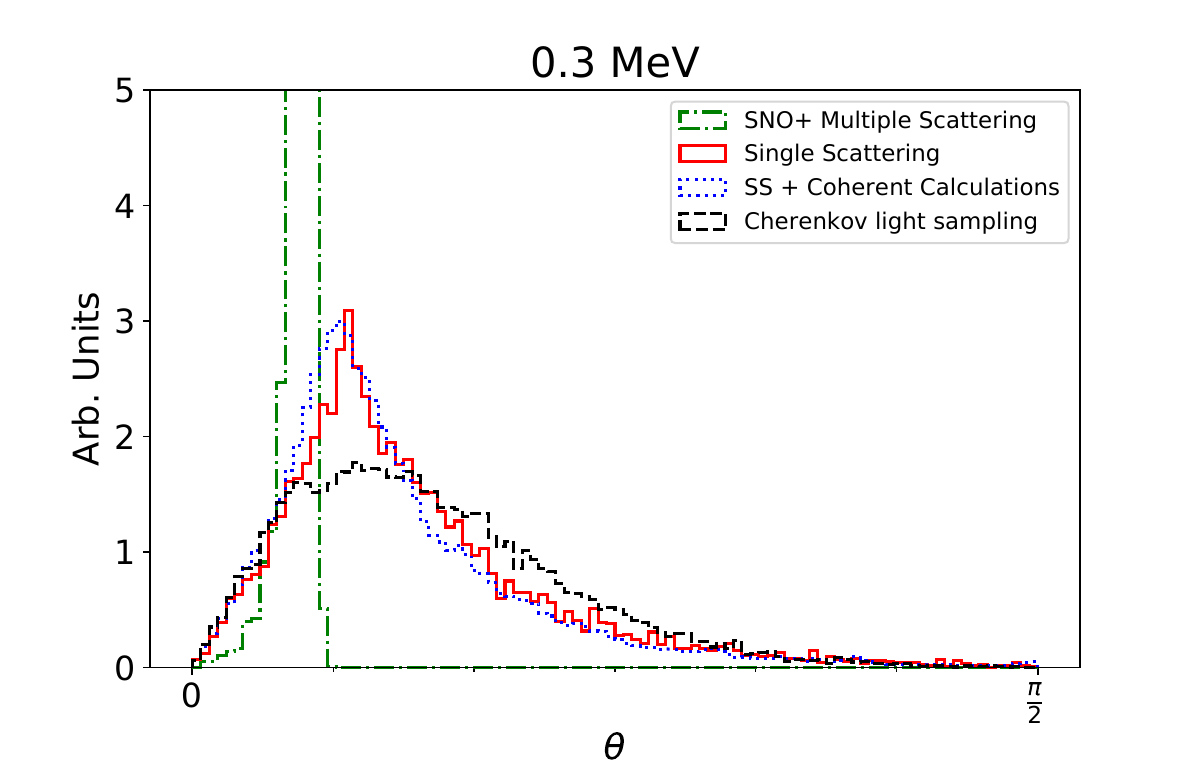}}
  \subfloat{\includegraphics[width=0.49\textwidth]{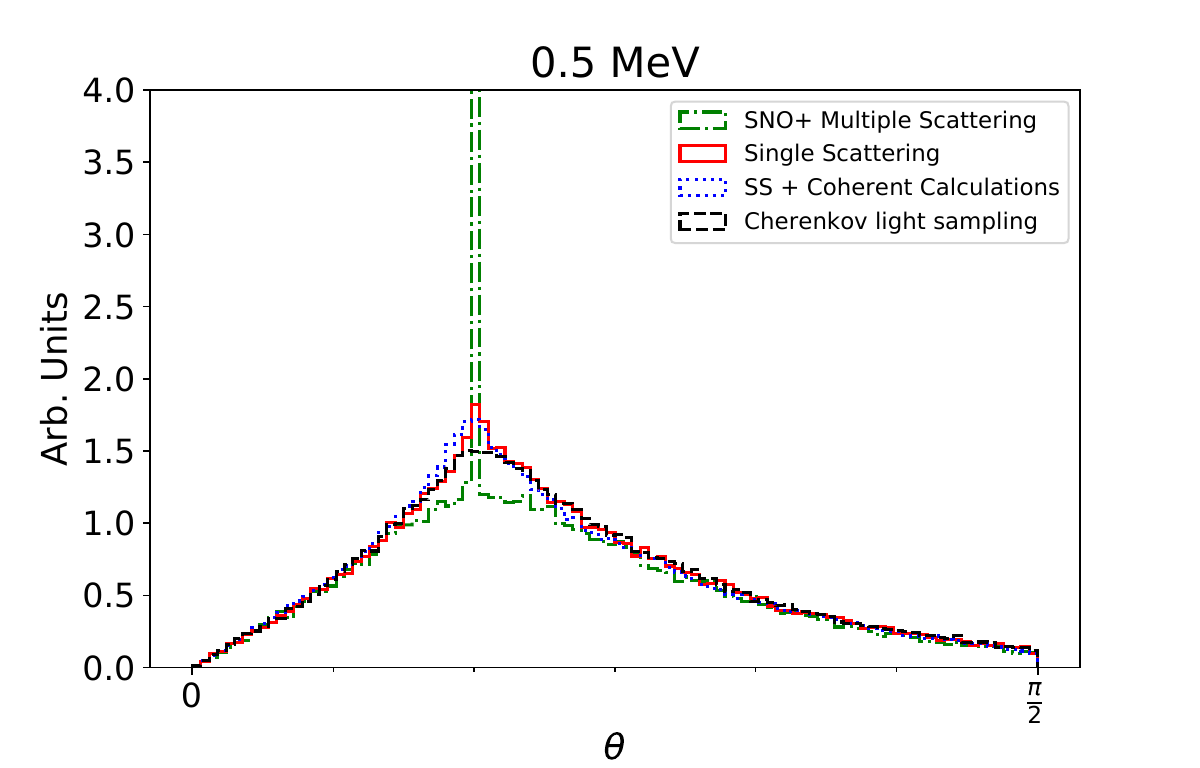}} \\
  \subfloat{\includegraphics[width=0.49\textwidth]{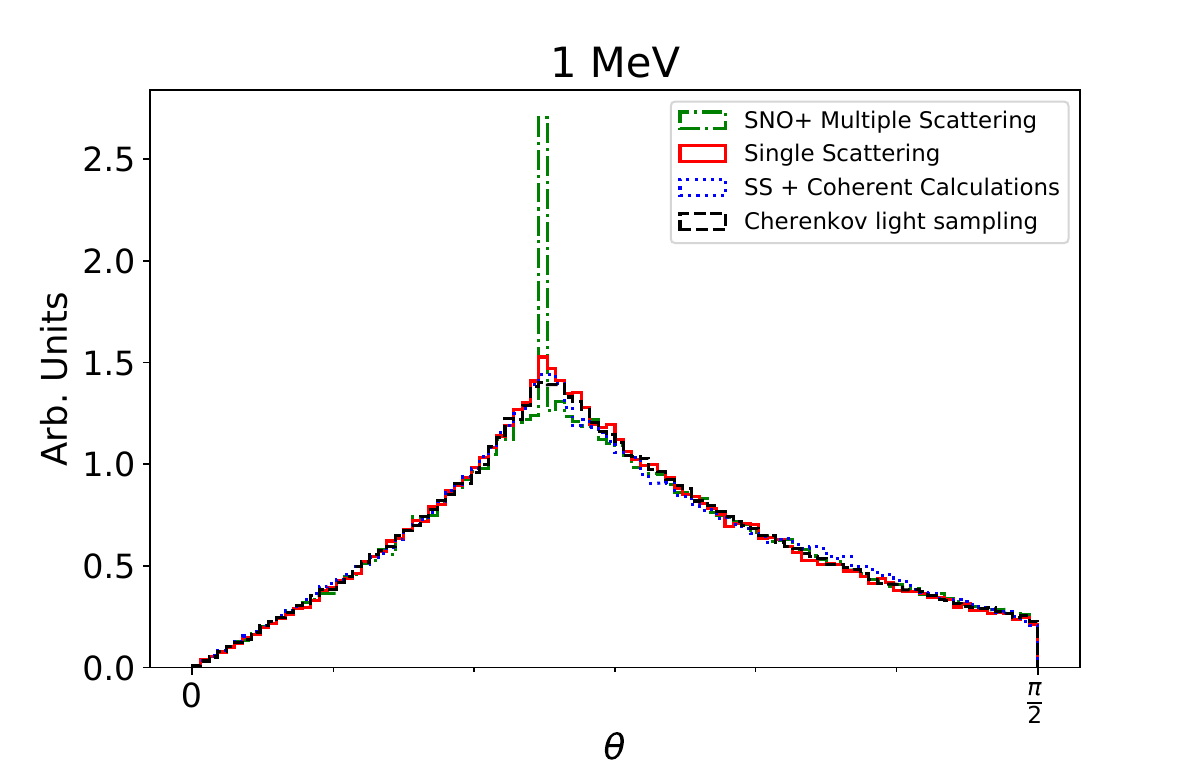}}
  \subfloat{\includegraphics[width=0.49\textwidth]{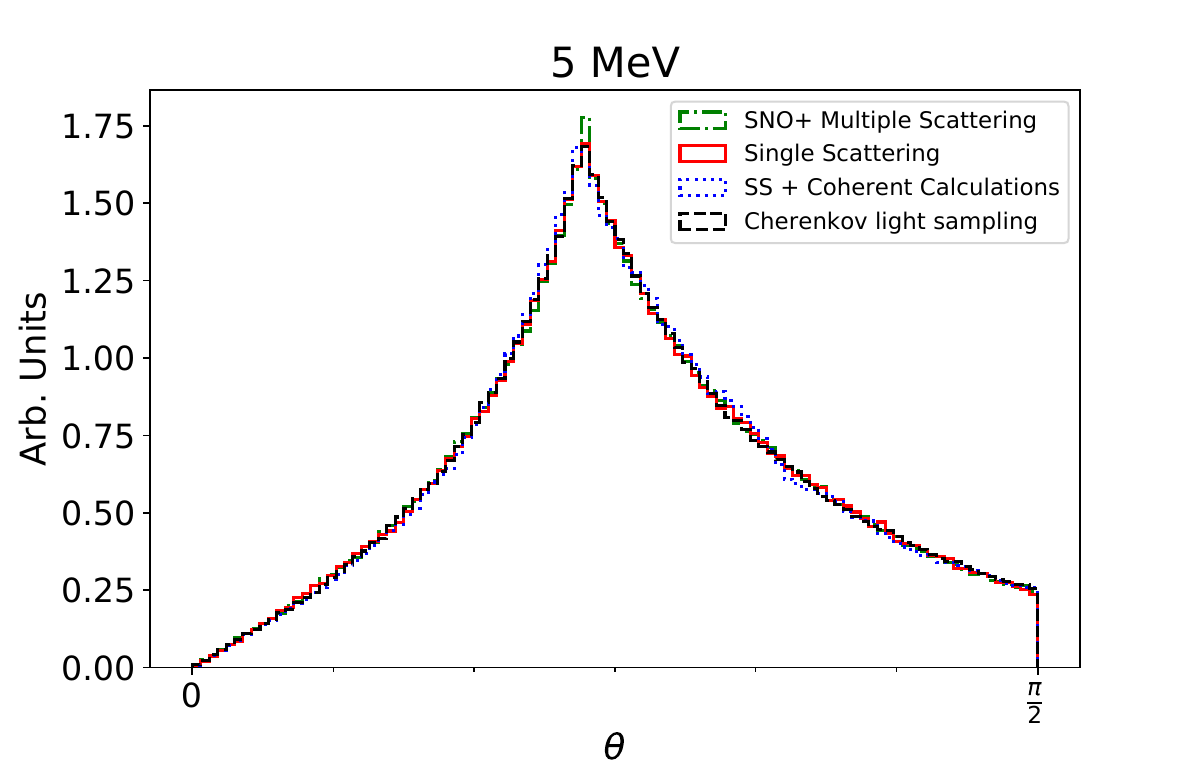}} \\
  \caption{Comparison of the angular distribution of Cherenkov light obtained from using the SNO+ default multiple scattering model, the Geant4 Single Scattering model, the single scattering model with coherent emission and the Cherenkov light sampling technique described in Sec.~\ref{ch:New_model}. Electrons are injected in the center of a sphere, pointing in the +z-direction. The angle of a photon is defined by their intersection with the sphere. Each panel corresponds to different electron initial energy.}
  \label{fig:SS_vs_MS_vs_coh}
\end{figure*}

A comparison of the angular distributions of the multiple scattering and single scattering models is shown in Fig.~\ref{fig:SS_vs_MS_vs_coh}. \Cref{tab:SSvsMS_2} summarizes the values of the three benchmark quantities. Both models produce a similar angular distribution when the electron energy reaches 5~MeV. However, for lower energies, the distributions differ significantly, with the multiple scattering model producing a sharper peak than the single scattering model. The single scattering model produces a cone with a larger half-angle. 

\begin{figure*}[!hbt]
    \centering
    \includegraphics[width=0.98\columnwidth]{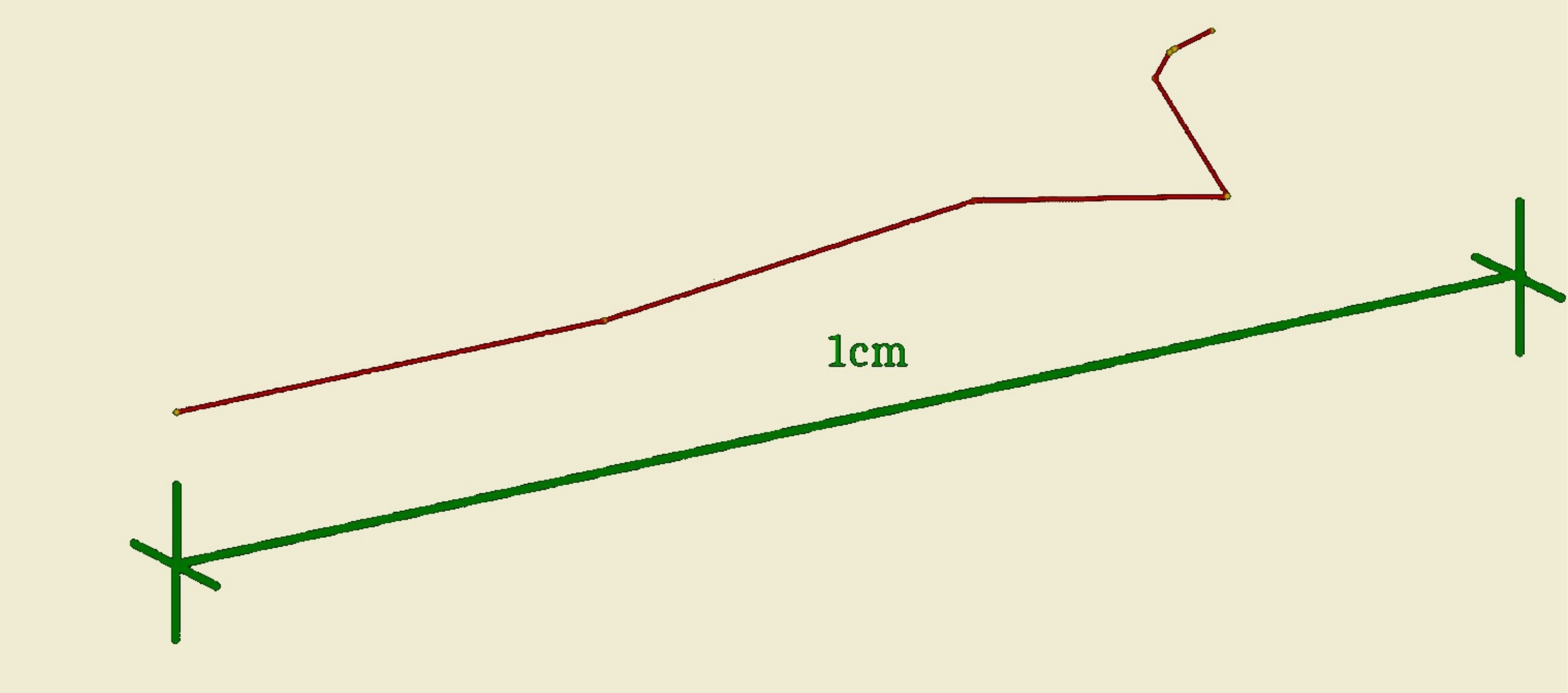}
    \includegraphics[width=0.98\columnwidth]{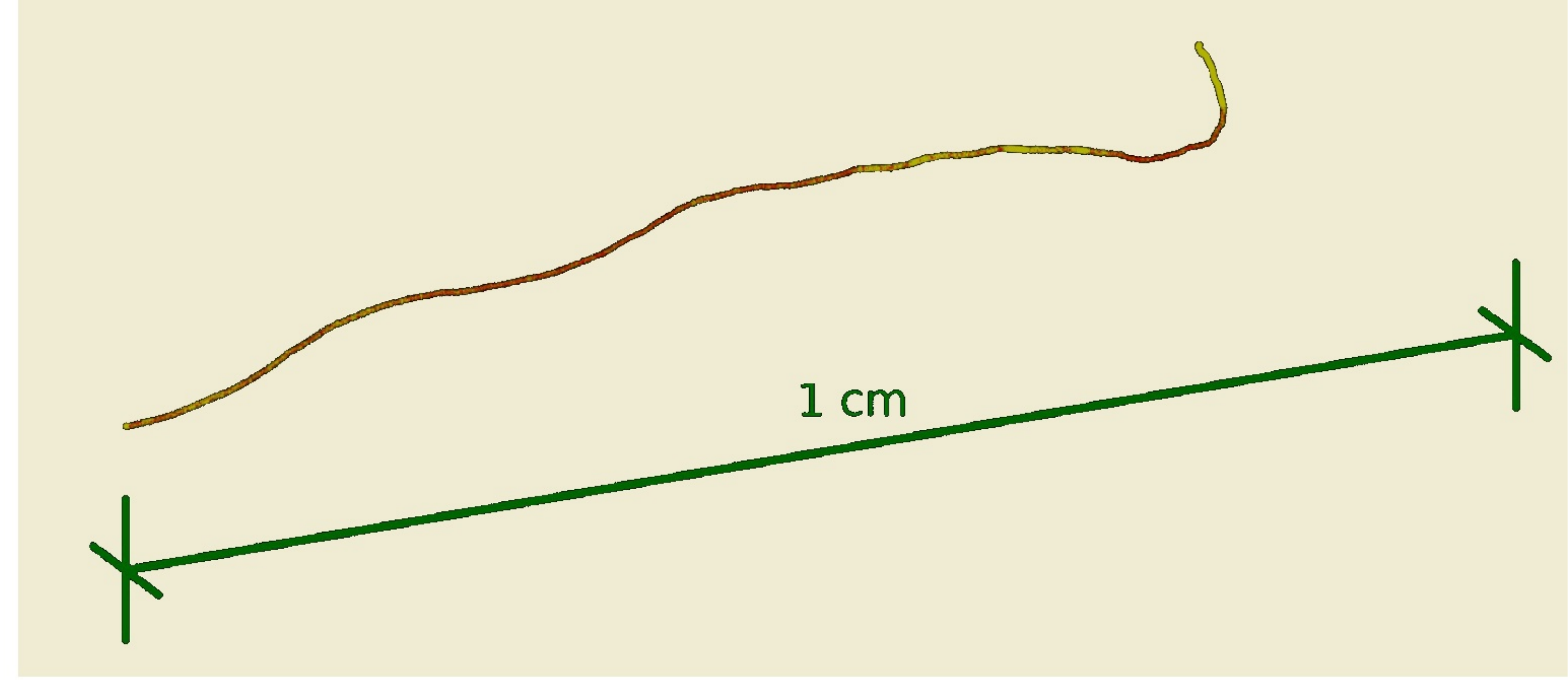}
    \caption{Geant4 simulation of a sample trajectory followed by 2~MeV electrons in water. Trajectories are shown in red. The left panel shows the result from using a multiple scattering model, while the panel on the right shows the trajectory produced by a single scattering model.}
    \label{fig:SS_vs_MS_traj}
\end{figure*}

The differences at low energies occur because the multiple scattering model produces large steps, of order 100~$\mu$s, for low energy electrons, emitting most of the light in the Cherenkov angle defined along the step. After a couple of steps, the electrons are below the Cherenkov threshold and cannot emit more light. However, electrons propagated with the single scattering model have comparatively short steps, of the order of a few $\mu$s, and most of these segments have enough energy to produce Cherenkov light. At 5~MeV and higher energies, even the multiple scattering model produces multiple steps before falling below the energy threshold for Cherenkov light production. As a result, both models produce very similar distributions. \Cref{fig:SS_vs_MS_traj} depicts these differences using a simulated 2~MeV electron in water, propagated with the multiple scattering and the single scattering models.

\begin{table*}
\centering
\vspace{20mm}
\begin{tabular}{l|c|c|c|c|c}
 Energy & Quantity &  Multiple Scattering & Single Scattering & Coherent & Sampling\\ \hline
\multirow{3}{4em}{0.3 MeV} 
& Peak position & 0.21 $\pm$ 0.03 & 0.31 $\pm$ 0.03 & 0.29 $\pm$ 0.03 & 0.38 $\pm$ 0.03 \\
& FWHM & 0.05$^{+0.06}_{-0.05}$ & 0.29 $\pm$ 0.06 & 0.25 $\pm$ 0.06 & 0.55 $\pm$ 0.06\\
& Tail fraction & 0.0001 $\pm$ 6e-5 & 0.0051 $\pm$ 0.0004 & 0.0083 $\pm$ 0.0006 & 0.0054 $\pm$ 0.0004 \\ \hline
\multirow{3}{4em}{0.5 MeV} 
& Peak position & 0.52 $\pm$ 0.03 & 0.57 $\pm$ 0.03 & 0.57 $\pm$ 0.03 & 0.59 $\pm$ 0.03 \\
& FWHM & 0.04$^{+0.06}_{-0.04}$ & 0.46 $\pm$ 0.06 & 0.46 $\pm$ 0.06 & 0.57 $\pm$ 0.06 \\
& Tail fraction & 0.036 $\pm$ 0.002 & 0.045 $\pm$ 0.002 & 0.040 $\pm$ 0.002 & 0.045 $\pm$ 0.002\\ \hline
\multirow{3}{4em}{1 MeV} 
& Peak position & 0.69 $\pm$ 0.03 & 0.71  $\pm$ 0.03 & 0.71 $\pm$ 0.03 & 0.72 $\pm$ 0.03 \\
& FWHM & 0.29 $\pm$ 0.06 & 0.55 $\pm$ 0.06 & 0.59 $\pm$ 0.06 & 0.62 $\pm$ 0.06 \\
& Tail fraction & 0.089 $\pm$ 0.001 & 0.081 $\pm$ 0.001 & 0.084 $\pm$ 0.001 & 0.090 $\pm$ 0.001 \\ \hline
\multirow{3}{4em}{5 MeV} 
& Peak position & 0.77 $\pm$ 0.03 & 0.77 $\pm$ 0.03 & 0.78 $\pm$ 0.03 & 0.76 $\pm$ 0.03\\
& FWHM & 0.46 $\pm$ 0.06 & 0.47 $\pm$ 0.06 & 0.50 $\pm$ 0.06 & 0.46 $\pm$ 0.06\\
& Tail fraction & 0.099 $\pm$ 0.001 & 0.092 $\pm$ 0.001 & 0.092 $\pm$ 0.001 & 0.102 $\pm$ 0.001 \\ 
\end{tabular}
\caption{\label{tab:SSvsMS_2} Comparisons of the distributions in Fig.~\ref{fig:SS_vs_MS_vs_coh} for different electron energies. For each model we compute the peak of the Cherenkov light distribution, the full width at half maximum (FWHM), as well as the fraction of photons emitted in the tails, with angle $> \pi / 2$. The comparatively large error on the tail fraction of low-energy multiple scattering models comes from the low number of photons that the model produces in this region (see Fig.~\ref{fig:SS_vs_MS_vs_coh}). }
\end{table*}

\subsection{Impact of interference effects in the Single Scattering model}
\label{ch:SSvsCoherentSS}
The most physically accurate Cherenkov simulation combines the single scattering model and the calculations in \Cref{ch:derivation}, which considers interference effects and coherence constraints. The simulation setup is the same as in the previous tests, except that Cherenkov photons are restricted to the single wavelength $\lambda = 400$~nm. The electron's position, time, and $\beta = v/c$ at every step are used to calculate the sum of the integrals $I_{\nu}$ from Eq.~\ref{eq:power_sum} and in this way obtain an angular distribution of Cherenkov light that includes interference effects. 

The interference term causes minimal deviations from the current simulation tools with the single scattering model, as shown in Fig.~\ref{fig:SS_vs_MS_vs_coh} and in~\Cref{tab:SSvsMS_2}. The results are in agreement with Ref.~\cite{BOWLER1996463}, where the authors state that there should not be any significant coherence violation of EM-radiation that forms Cherenkov light due to electron scattering in water. 

\FloatBarrier

\section{Cherenkov light sampling: a fast implementation of the improved Cherenkov light model}\label{ch:New_model}

The single scattering model is significantly more computationally expensive than multiple scattering models; the simulations in~\Cref{ch:MSvsSS} differed in computational time by a factor of 7. We developed and tested an algorithm we call ``Cherenkov light sampling'' that corrects the multiple scattering model and approximates the angular distribution of single scattering models with almost negligible impact on computational speed. 

For each medium, the single scattering model is used to generate the probability distributions for single events scatters, $\theta_{ss}(E_e)$ and the free mean path of the electron $\lambda_{ss}(E_e)$ as a function of the electron's energy. We binned these distributions in steps of 50 keV for $\theta_{ss}(E_e)$ and steps of 100 keV for $\lambda_{ss}(E_e)$, and then interpolated with a smooth function to have access to arbitrary energies. We chose the steps to be smaller than the electron energies being tested, and we confirmed that modifying their values did not impact the results.

\begin{figure}[!tb]
    \centering
    \begin{tikzpicture}[node distance=2cm, scale=1, every node/.style={scale=0.75}]
    \node (start) [startstop] {Start};
    \node (in1) [io, below of=start, yshift=0.5cm] {Obtain $L_i$, $N_i$, $E_i$, $E_{i+1}$, and $\vec{p}_i$ by multiple scattering model };
    \node (pro1) [process, below of=in1] {$L_p$ = $\frac{L_i}{N_i}$ \\
    $\Delta E = \frac{E_i - E_{i+1}}{N_i}$ \\
    $E_{ij} = E_i + \frac{1}{2} \Delta E$ \\
    j = 0
    };
    \node (pro2) [process, below of=pro1, yshift=-0.5cm] {
    $E_{ij} = E_{ij} - \Delta E$ \\
    $N_{\theta} = \frac{L_p}{\lambda_{ss}(E_{ij})}$ \\
    jj = 0};
    \node (pro3) [process, below of=pro2] {
    Sample $\Delta \theta$ from $\theta_{ss}(E_{ij})$ \\};
    
    \node (pro7) [rectangle,
    minimum width=1cm, 
    minimum height=1cm, 
    text centered, 
    draw=black,
    minimum width=1.6cm,
    fill=orange!30, xshift=-1.7cm, left of=pro2] { j = j + 1};
    
    \node (pro4) [process, below of=pro3] {
    Sample $\Delta \varphi$ from $[0; 2\pi]$ interval. \\
    $\vec{p}_i$.theta = $\vec{p}_i$.theta + $\Delta \theta$ \\
    Rotate $\vec{p}_i$ for $\Delta \varphi$};
    
    \node (pro5) [rectangle,
    minimum width=1cm, 
    minimum height=1cm, 
    text centered, 
    draw=black, 
    minimum width=1.6cm, 
    fill=orange!30, xshift=-1.7cm, left of=pro4] { jj = jj + 1};
    
    \node (dec1) [decision, below of=pro4, yshift=-0.3cm] {Is jj < $N_{\theta}$ ?};
    
    \node (pro6) [process, below of=dec1, yshift=-0.3cm] {Emit a photon w.r.t. $\vec{p}_i$};
    
    \node (dec2) [decision, below of=pro6] {Is j < $N_i$ ?};
    
    \node (stop) [startstop, below of=dec2, yshift=-0.3cm] {Stop};
    
    \draw [arrow] (start) -- (in1);
    \draw [arrow] (in1) -- (pro1);
    \draw [arrow] (pro1) -- (pro2);
    \draw [arrow] (pro2) -- (pro3);
    \draw [arrow] (pro3) -- (pro4);
    \draw [arrow] (pro4) -- (dec1);
    \draw [arrow] (dec1.west) -- ++(-1.65,0) -- ++(0,1) -- node[xshift=3cm,yshift=-1.3cm, text width=2.5cm] {Yes} (pro5);
    \draw [arrow] (pro5) |- node {} (pro3);
    \draw [arrow] (dec1) -- node[xshift=0.5cm, yshift=0.2cm] {No} (pro6);
    \draw [arrow] (dec2.west) -- ++(-3.2,0) -- ++(0,7.95) -- ++ (0.1,0) -- node[xshift=3.8cm,yshift=-10.3cm, text width=2.5cm] {Yes} (pro7);
    \draw [arrow] (pro7) -- (pro2);
    \draw [arrow] (pro6) -- (dec2);
    \draw [arrow] (dec2) -- node[xshift=0.5cm, yshift=0.2cm] {No} (stop);
    
    \end{tikzpicture}

    \caption{Flow chart of the Cherenkov light sampling algorithm. Here $L_i$ is the length of the $i-$th step of the propagated electron with a mulitiple scattering model, $N_i$ is number of photons for that step, $E_i$ and $E_{i+1}$ are the kinetic energies of the electron at the beginning and at the end of the segment and $\vec{p}_i$ is the direction of the segment.}
    \label{fig:flow_chart}
\end{figure}
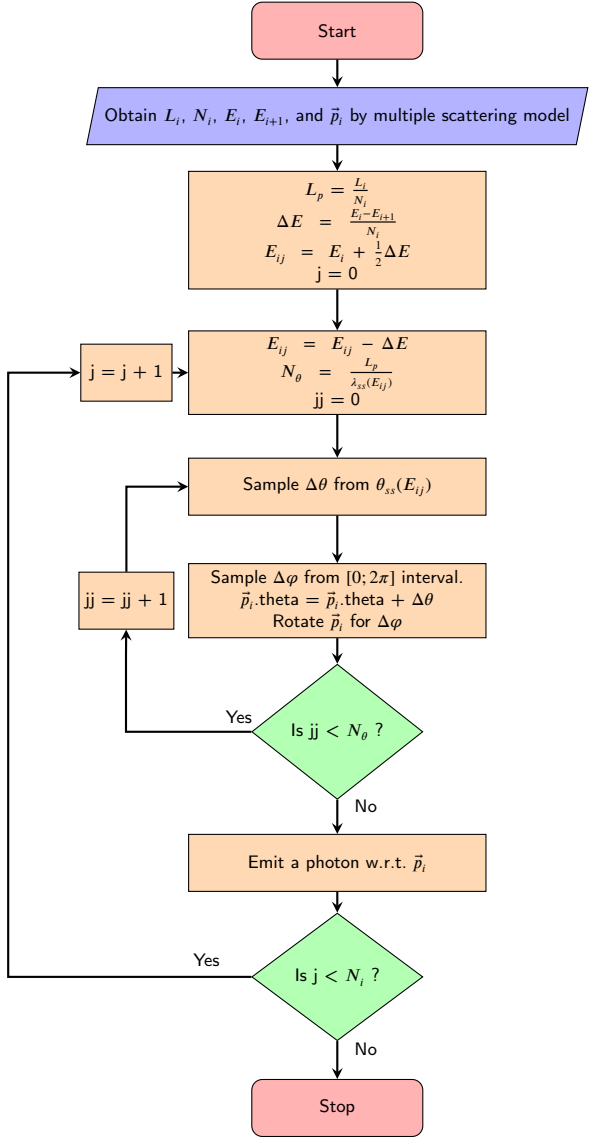

 To simulate Cherenkov photons, we then use a multiple-scattering model, which generates a step length and particle direction, and determines the number of Cherenkov photons. A more detailed trajectory is generated with a length sampled from $\lambda_{ss}$ and scattering angle from $\theta_{ss}(E_e).$ Cherenkov photons are generated uniformly along the detailed trajectory. The algorithm is explained in Fig.~\ref{fig:flow_chart}.

We assessed the accuracy with which the Cherenkov light sampling reproduces the result of the single scattering model using electrons in water as in \Cref{ch:MSvsSS}. \Cref{fig:SS_vs_MS_vs_coh} and \Cref{tab:SSvsMS_2} summarize the distributions. Cherenkov light sampling essentially reproduces the results from the single scattering model, although minor differences remain, particularly at 0.3~MeV. We observe a 7\% increase of computation time compared to the multiple scattering model.

\begin{figure}[!tb]
    \centering
    \includegraphics[width=0.5\textwidth]{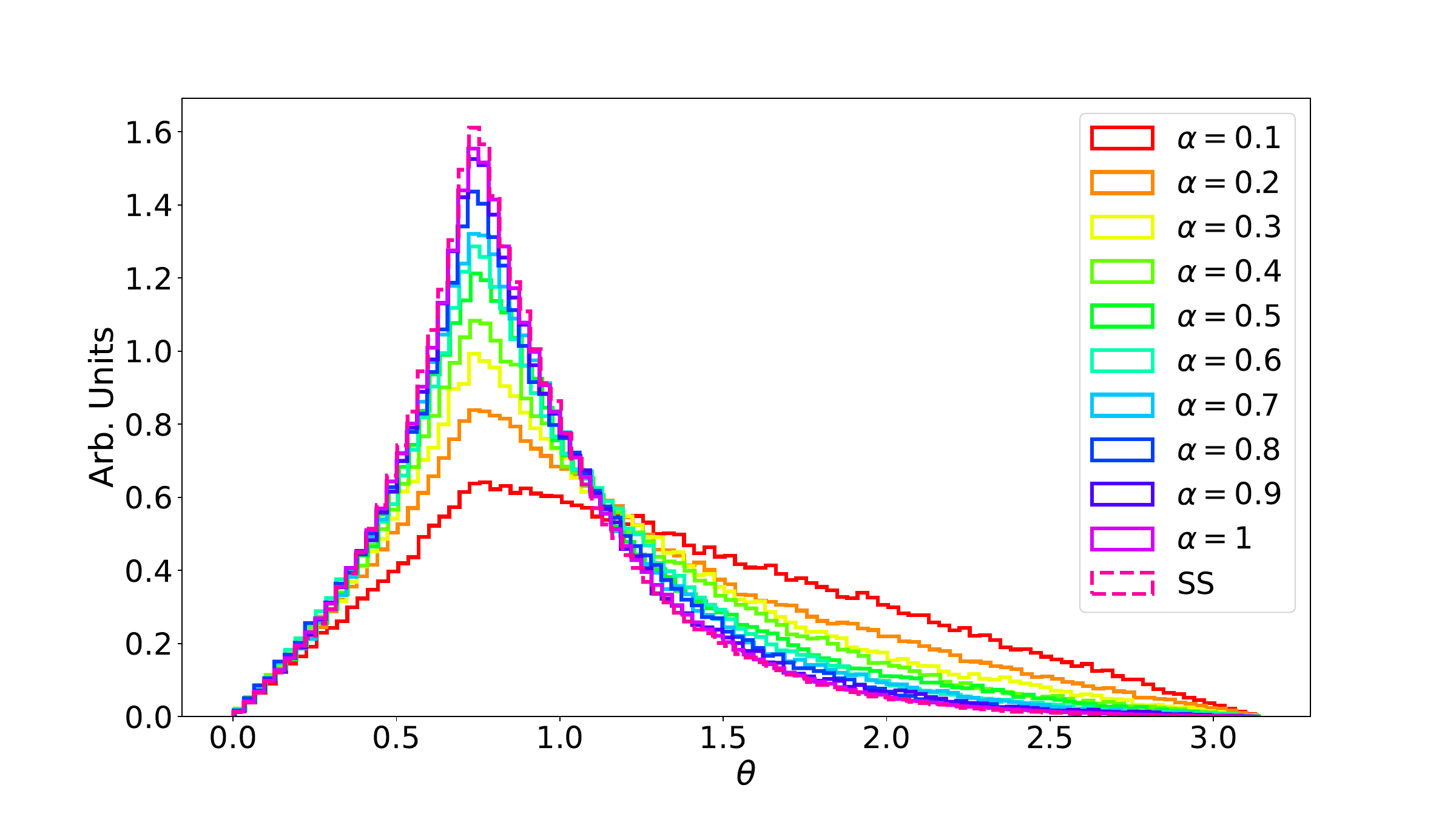}
    \caption{Comparison of the Cherenkov light distributions produced by the Cherenkov light sampling method using different $\alpha$ values. The simulation is the same as in Fig.~\ref{fig:SS_vs_MS_vs_coh} with an electron initial energy of 1~MeV.}
    \label{fig:alpha_smearing}
\end{figure}

We can compensate for the some of the differences between the Cherenkov sampling and the single scattering model and for differences due to neglecting the interference term by introducing a parameter $\alpha$ that scales the mean free path of an electron as $\lambda^{\prime}_{ss}(E) = \alpha \lambda_{ss}(E)$. A value smaller than one results in an additional smearing of the Cherenkov light angular distribution, as shown in Fig.~\ref{fig:alpha_smearing}. 

We ran simulations of the Cherenkov sampling method using different values of $\alpha$, and found that the best agreement with the results from the single scattering model were obtained for values of $\alpha \sim 0.6$. This indicates that the approximations in the method do not fully capture the effects from the single scattering model and require further smearing of the photon injection angle. In the analysis that follows we test the method on experimental data, and we fit for the value of $\alpha$ that describes the data best.

\FloatBarrier

\subsection{Cherenkov light sampling with SNO+ data}\label{ch:Tuning}
    
SNO+ is a large-scale underground neutrino detector~\cite{SNO:2021xpa} consisting of a spherical acrylic vessel of 6~m radius, submerged in water and surrounded by a structure of 9,394 photomultiplier tubes (PMTs). The PMTs provide about 50\% photo coverage for events inside the vessel. The data discussed in this section comes from the water phase of the experiment.

In the water phase of SNO+, events are detected if they produce Cherenkov light. A single electron produces a well defined ring-like pattern on the PMTs with an average number of 9~PMT hits per MeV. The angular distribution has been used to distinguish between different single electrons, gammas, and gamma cascades~\cite{SNO:2018ydj, SNO:2022trz}. SNO+ characterizes the isotropy of the light using parameters derived from the angular distribution of the PMTs, initially described in Ref.~\cite{dunmore_2004}.

\begin{figure}[!tb]
    \centering
    \includegraphics[width=0.45\textwidth]{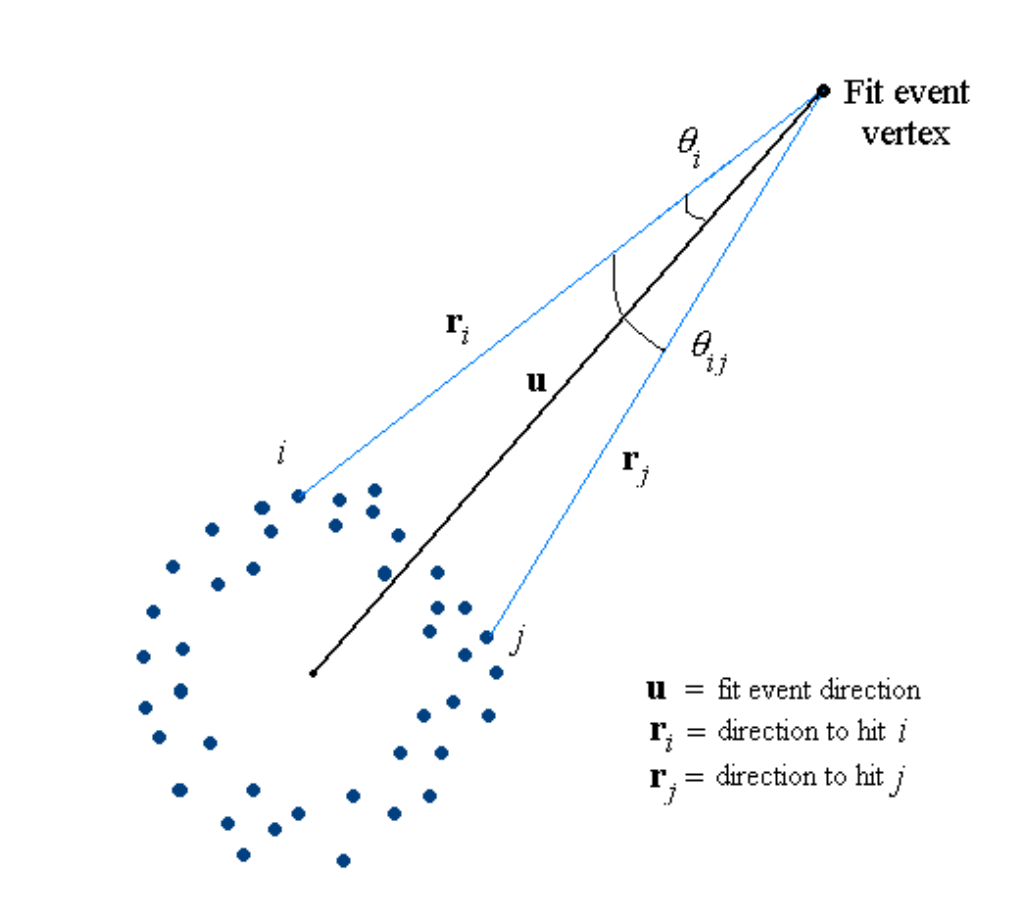}
    \caption{The $\theta_{ij}$ angles within a Cherenkov ring event used to calculate $\beta_{14}$. Reproduced from \cite{dunmore_2004}. }
    \label{fig:event_scheme}
\end{figure}

\Cref{fig:event_scheme} shows a schematic representation of PMT hits by Cherenkov light, where the ring structure is clearly visible. For a particle at vertex position $\textbf{v}$
travelling along unit vector $\hat{u}$,
and with PMTs located at $\textbf{p}_i, \textbf{r}_i=\textbf{p}_i-\textbf{v}.$ We define
$\theta_{ij}$ by $\cos{\theta_{ij}} = \frac{\textbf{r}_i \textbf{r}_j}{|\textbf{r}_i| |\textbf{r}_j|}$ and
\begin{equation}
    \beta_l = \frac{2}{N(N-1)} \left[ \sum^{N-1}_{i=1} \sum^N_{j = i+1} P_l(\cos{\theta_{ij}}) \right],
\end{equation}
where $N$ is number of hits in the event and $P_l$ is a Legendre polynomial of degree $l$.
For SNO~\cite{SNO:1999crp}, the predecessor of SNO+, it was determined\cite{dunmore_2004} that the best differentiation was achieved using $\beta_{14} = \beta_1 + 4 \beta_4$. The value of $\beta_{14}$ is anticorrelated with the isotropy of the light.

In the water phase of SNO+ a tension in $\beta_{14}$ between data and simulation~\cite{SNO:2018ydj} was observed. Since the parameter contains the most information on the angular distribution of Cherenkov light, we will use it to fit the $\alpha$ parameter of the Cherenkov light sampling model to calibration data, and compare the results obtained. Moreover, since the SNO+ simulation for the water phase uses Geant4 for propagation of particles, we can directly apply the Cherenkov light sampling model to conduct our tests.

The calibration data chosen for the comparisons come from the deployment of radioactive sources in the center of SNO+. The two calibration devices used are a $^{16}$N source~\cite{N16} and an AmBe source~\cite{AmBe}, which provide nearly monoenergetic 6.13~MeV, 4.4~MeV, and 2.2~MeV $\gamma$-rays. We run simulations with different $\alpha$ in a range from 0.1 to 1 for the $^{16}$N and AmBe sources placed in the middle of the acrylic vessel, using the corresponding detector conditions for each run. As a result we obtain $\beta_{14}$ distributions as a function of $\alpha$ for three different energies of $\gamma$-rays. 

\begin{figure}[!b]
    \centering
    \includegraphics[width=1\columnwidth]{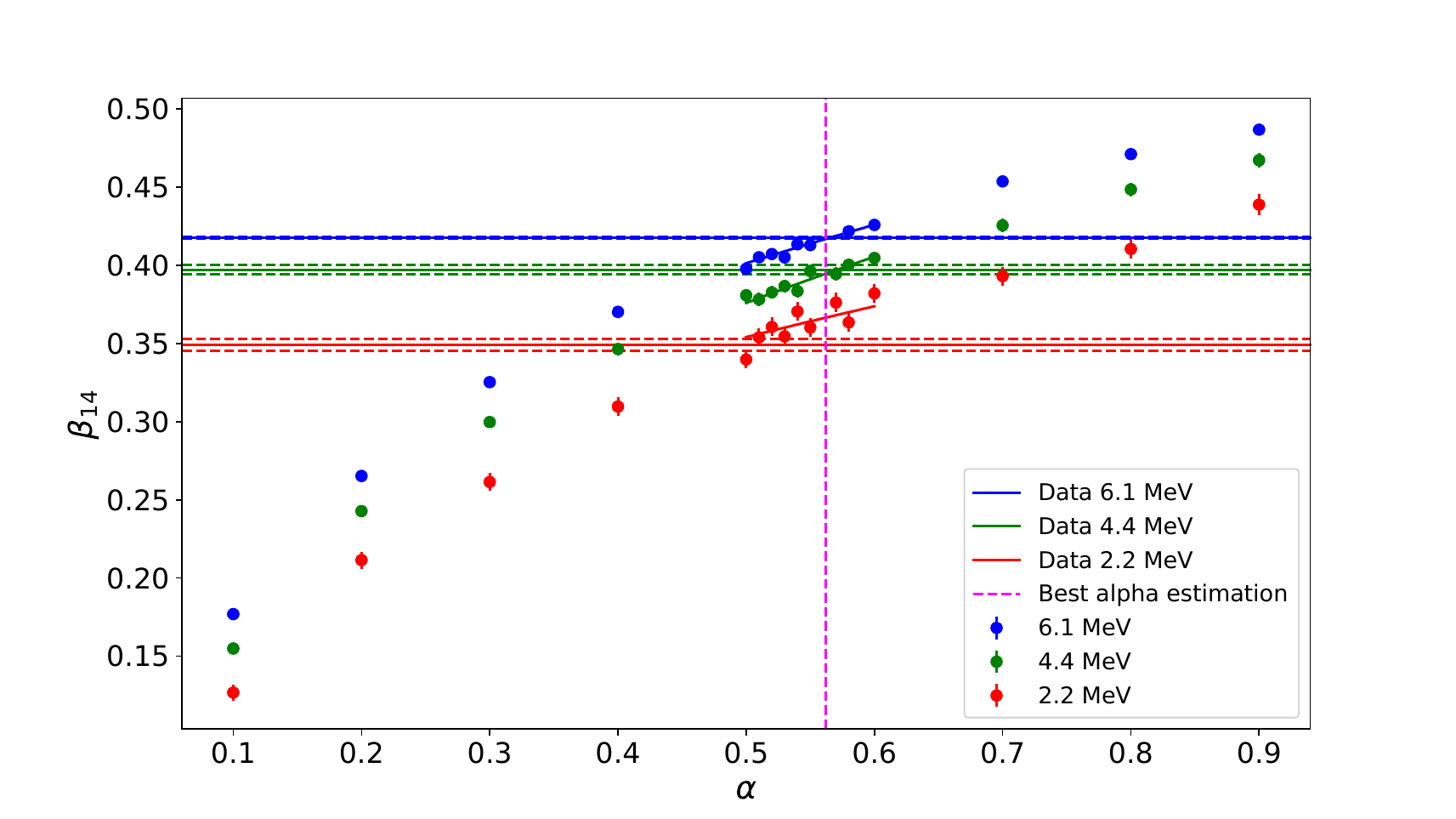}
    \caption{Comparison of $\beta_{14}$ mean values obtained from simulations as a function of the parameter $\alpha$ together with the curve that fits them best, described in the text. The $\beta_{14}$ mean values from calibration sources with $\gamma$ - rays of energies 2.2 MeV, 4.4 MeV, and 6.1 MeV are also shown. The optimal value for $\alpha$ is shown by the pink dashed line.}
    \label{fig:beta_alpha_func}
\end{figure}

\Cref{fig:beta_alpha_func} shows the values of the mean of a Gaussian fit of these $\beta_{14}$ distributions as well as the values from the calibration data. While the overall dependence of the mean of $\beta_{14}$ on $\alpha$ is clear, the results show a significant level of noise when making small changes in $\alpha$. This is more obvious for the 2.2~MeV gamma. For this reason, we opted for fitting the relationship between $\beta_{14}$ and $\alpha$ as $\beta_{14}(\alpha) = a \sqrt{\alpha} + b$ for the range $\alpha=[0.5,0.6]$. To find an optimal value of $\alpha$, we then minimize the function
\begin{equation}
    \label{eq:minim_chi2}
    \chi^2(\alpha) = \sum_i \frac{(\beta_{14_i}(\alpha) - \beta^{data}_{14_i})^2}{\sigma^2_{data_i} + \sigma^2_{MC_i} (\alpha)},
\end{equation}
where $i$ = 6.1~MeV, 2.2~MeV, 4.4~MeV, and $\sigma^2_{data}$ and $\sigma^2_{fit}$ are errors of $\beta_{14}$ data and MC respectively. The result of the fit is $\alpha = 0.556 \pm 0.005$, which is close to the expectation from simulations of 0.6. To verify this result, we produced new simulation for the calibration sources using the fit value of $\alpha$, and compare the expectation of the full $\beta_{14}$ distribution, as well as angular distribution of the Cherenkov light observed.

\begin{table*}
\centering
\begin{tabular}{l|c|c|c}
Calibration source & $^{16}$N & AmBe$_{4.4\mathrm{MeV}}$ & AmBe$_{2.2\mathrm{MeV}}$ \\ \hline
Data $\beta_{14}$ mean & 0.4176 $\pm$ 4e-4 & 0.397 $\pm$ 2e-3 & 0.350 $\pm$ 3e-3 \\
Default MC $\beta_{14}$ mean & 0.4414 $\pm$ 5e-4 & 0.422 $\pm$ 2e-3 & 0.387 $\pm$ 3e-3 \\
Sampling $\beta_{14}$ mean & 0.4177 $\pm$ 7e-4 & 0.399 $\pm$ 2e-3 & 0.367 $\pm$ 3e-3 \\ \hline
Data $\beta_{14}$ $\sigma$ & 0.1727 $\pm$ 3e-4 & 0.201 $\pm$ 2e-3 & 0.290 $\pm$ 3e-3 \\
Default MC $\beta_{14}$ $\sigma$ &  0.1760 $\pm$ 3e-4 & 0.209 $\pm$ 1e-3 & 0.297 $\pm$ 2e-3 \\
Sampling MC $\beta_{14}$ $\sigma$ & 0.1714 $\pm$ 5e-4 & 0.207 $\pm$ 2e-3 & 0.292 $\pm$ 3e-3 \\ 
\hline
Default MC $\beta_{14}$\;\;$\chi^2$/d.o.f. & 1085.6/80 & 141.6/80 & 117.6/80 \\
Sampling MC $\beta_{14}$\;\;$\chi^2$/d.o.f. & 90.4/80 & 69.6/80 & 78.4/80 \\
\hline
Default MC $\cos\theta_{PMT}$\;\;$\chi^2$/d.o.f.& 8706/200 & 9312/200 & 12374/200 \\
Sampling MC $\cos\theta_{PMT}$ \;\;$\chi^2$/d.o.f.& 1772/200 & 7523/200 & 10763/200 \\
\end{tabular}
\caption{\label{tab:beta14_results} Comparison of variables that encode information of the angular distribution of Cherenkov light for data and simulation. The mean and $\sigma$ of $\beta_{14}$ from the Gaussian fits are shown, as well as the $\chi^2$ over degrees of freedom for $\beta_{14}$ and $\cos\theta_{PMT}$.}
\end{table*}

\paragraph{$\beta_{14}$ -- }

\begin{figure}[!b]
    \centering
    \includegraphics[width=1\columnwidth]{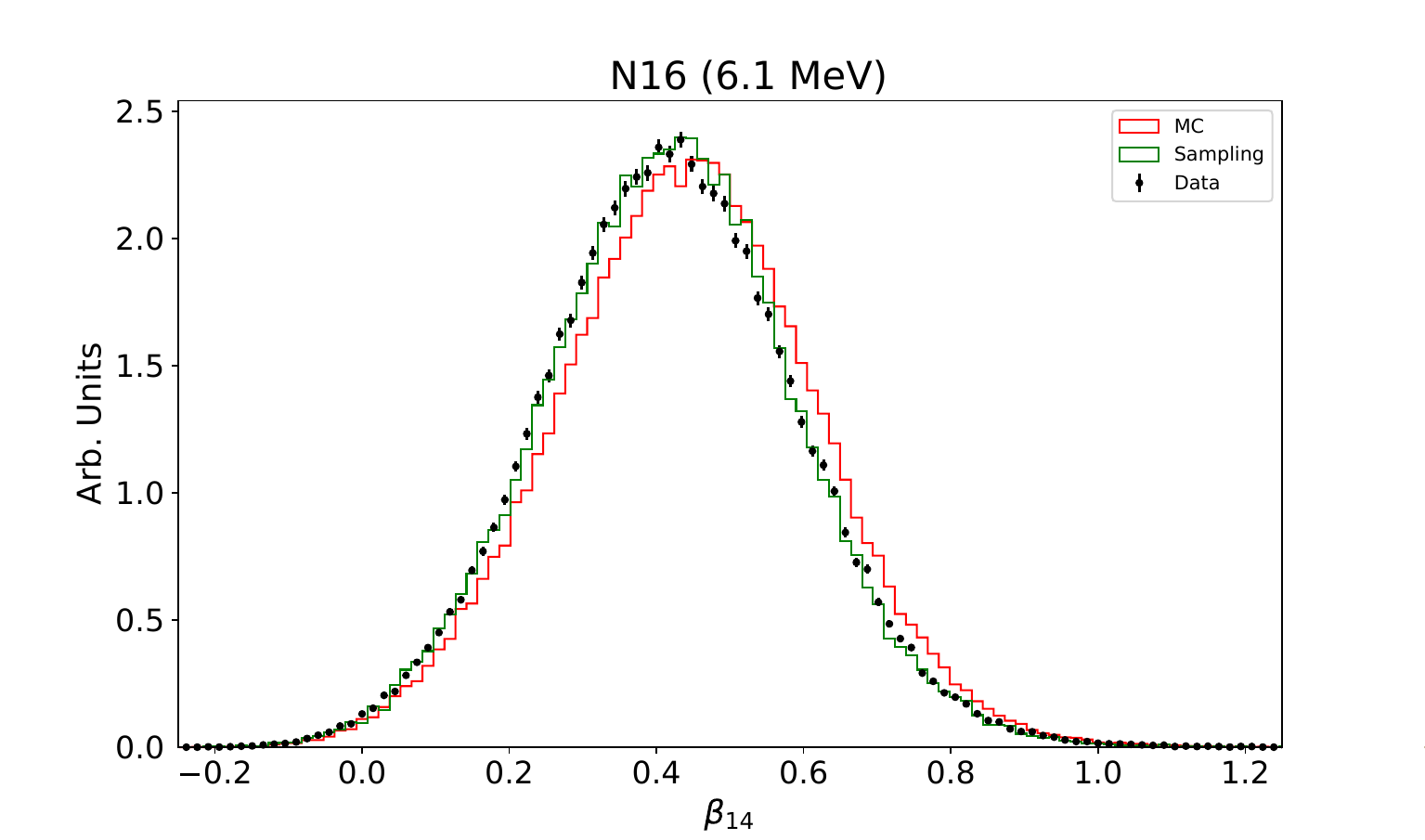}
    \caption{Comparison of the $\beta_{14}$ distribution of the of $^{16}$N (6.1~MeV) for data, the default SNO+ MC and simulation produced using the Cherenkov light sampling method.}
    \label{fig:beta14_N16}
\end{figure}

The complete $\beta_{14}$ distributions for the three $\gamma$-ray energies considered are shown in Figs.~\ref{fig:beta14_N16} and~\ref{fig:beta14_AmBe}. The improvement in agreement can be seen directly in the figures, particularly for the 6.1~MeV gammas. \Cref{tab:beta14_results} quantifies this agreement, and also summarizes and compares the properties of the distributions. The Cherenkov sampling model explains the mean of the distributions by design, but it also describes their width better than the default simulation. The agreement in the shape of the distribution is highlighted by the $\chi^2$, which shows significant improvement for all three $\gamma$-ray energies.

\begin{figure}[!bt]
    \centering
    \includegraphics[width=1\columnwidth]{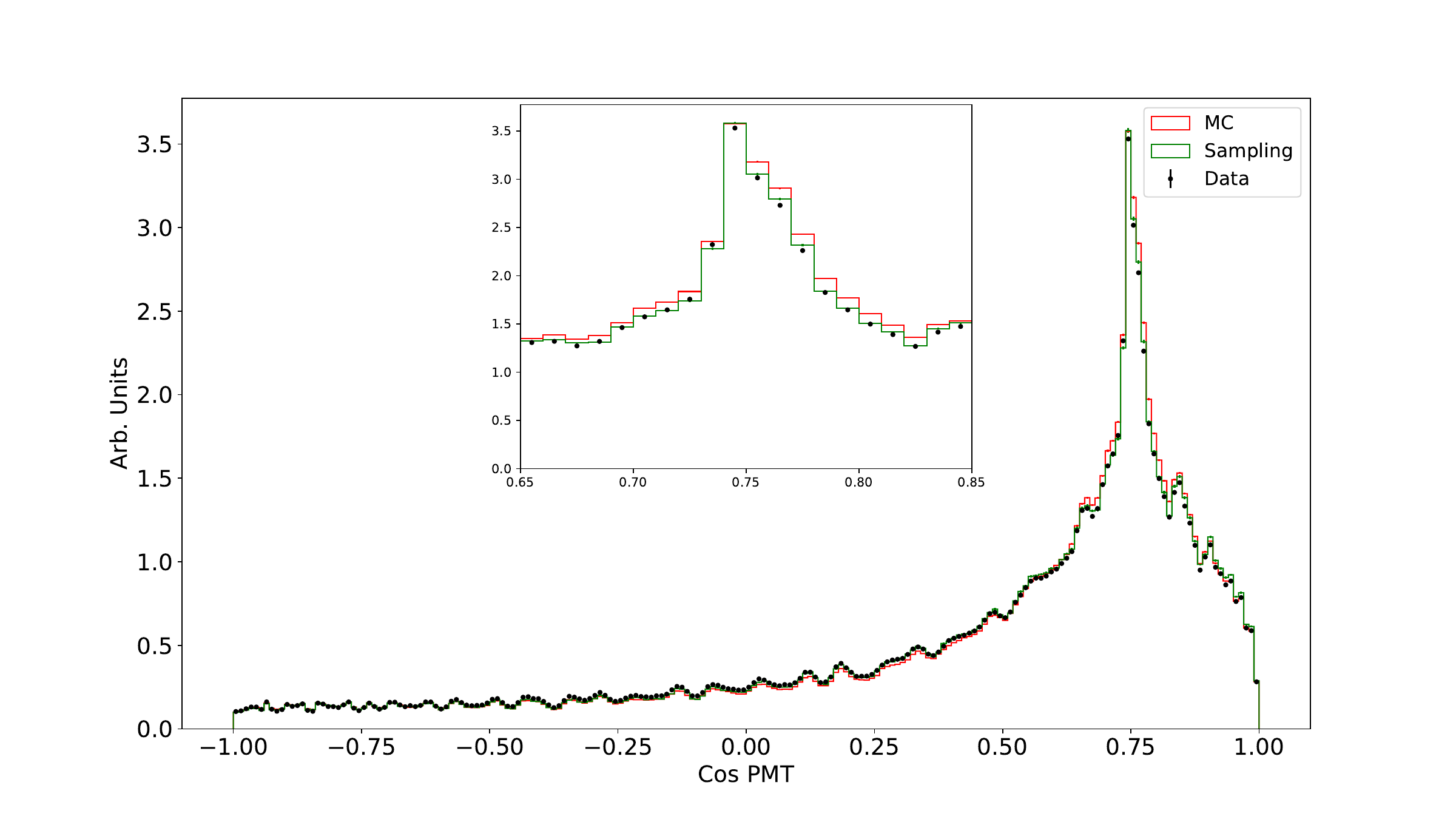}
    \caption{Comparison of the $\cos\theta_{PMT}$ distribution of the $^{16}$N (6.1~MeV) for data, the default SNO+ MC and simulation produced using the Cherenkov light sampling method.}
    \label{fig:CosPMT_N16}
\end{figure}

\paragraph{$\cos\theta_{PMT}$} -- The angular distribution of the Cherenkov light is quantified by $\cos{\theta}_{PMT}$, obtained as
\begin{equation}
    \label{eq:cosPMT}
    \cos{\theta}_{PMT} = \vec{u}_{fit} \cdot \frac{\vec{x}_{PMT} - \vec{x}_{fit}}{\left|\vec{x}_{PMT} - \vec{x}_{fit}\right|},
\end{equation}
where $\vec{x}_{PMT}$ is the coordinate of the triggered PMT, $\vec{x}_{fit}$ is the reconstructed position of the electron observed, and $\vec{u}_{fit}$ is its reconstructed direction. This variable is similar to Fig.~\ref{fig:SS_vs_MS_vs_coh}, now including the detector response: geometry, PMTs response and photon transport.
\Cref{fig:CosPMT_N16} shows the $\cos{\theta}_{PMT}$ distribution for the $^{16}$N (6.1~MeV) source, while the distributions for the AmBe source are shown in Appendix. A quantitative description of the distributions and their agreement is given in~\Cref{tab:beta14_results}. Here, the sampling model also shows a significant improvement with respect to the default simulation. The improvements are most noticeable for the $^{16}$N calibration source, despite the similarity in the distributions of sampling and multiple scattering on Fig.~\ref{fig:SS_vs_MS_vs_coh}.

Overall, the Cherenkov sampling method shows an improved agreement between the MC and data in the two variables that encode information about the angular emission of the light. Interestingly, these changes do not modify the behavior of other variables typically used in analyses, which mainly depend on timing and number of detected photons.

\FloatBarrier

\section{Conclusion}
\label{sec:conclusion}

In this paper, we investigated the relevance of the interference effects and the electron propagation method for simulation of Cherenkov light at a few MeVs. We demonstrated that the choice of electron propagation method has a considerable impact on the overall distribution of Cherenkov light below 6~MeV, while interference effects play a much smaller role. Consequently, we developed an improved Cherenkov light simulation model specifically tailored for MeV electrons propagating in a uniform medium. This model is based on the single scattering model and provides an improved level of accuracy while maintaining computational efficiency similar to previous methods. The model requires a one-time calculation of electron scattering in the relevant medium, and has a single parameter that can be tuned to relevant experimental data.
We implemented the method in the Geant4 package and used it on SNO+ calibration data. After tuning the model, we are able to significantly improve the description of variables that depend on the angular distribution of Cherenkov light.
Our findings might be relevant for experimental setups that use the angular distribution of Cherenkov light, in particular new projects that will have improved sensitivity to these effects, such as Hyper-Kamiokande~\cite{Abe:2011ts}, currently under construction, or the planned THEIA detector~\cite{Theia:2019non}.

\section*{Acknowledgements}
This research was undertaken thanks in part to funding from the Natural Sciences and Engineering Research Council of Canada (NSERC), the Canada First Research Excellence Fund through the Arthur B. McDonald Canadian Astroparticle Physics Research Institute, WestGrid and the Digital Research Alliance of Canada\footnote{www.alliancecan.ca} (formerly known as Compute Canada). The authors extend their thanks to the SNO+ Collaboration for access to the calibration data that spurred the questions addressed here, allowing us to study the Cherenkov light sampling method, and to Eric V\'azquez J\'auregui and Logan Lebanowski for reviewing the manuscript.

\printcredits

\bibliographystyle{model1-num-names}

\bibliography{cas-refs}


\appendix
\section{Appendix}

\begin{figure*}[!b]
    \centering
    \subfloat{\includegraphics[width=0.46\textwidth]{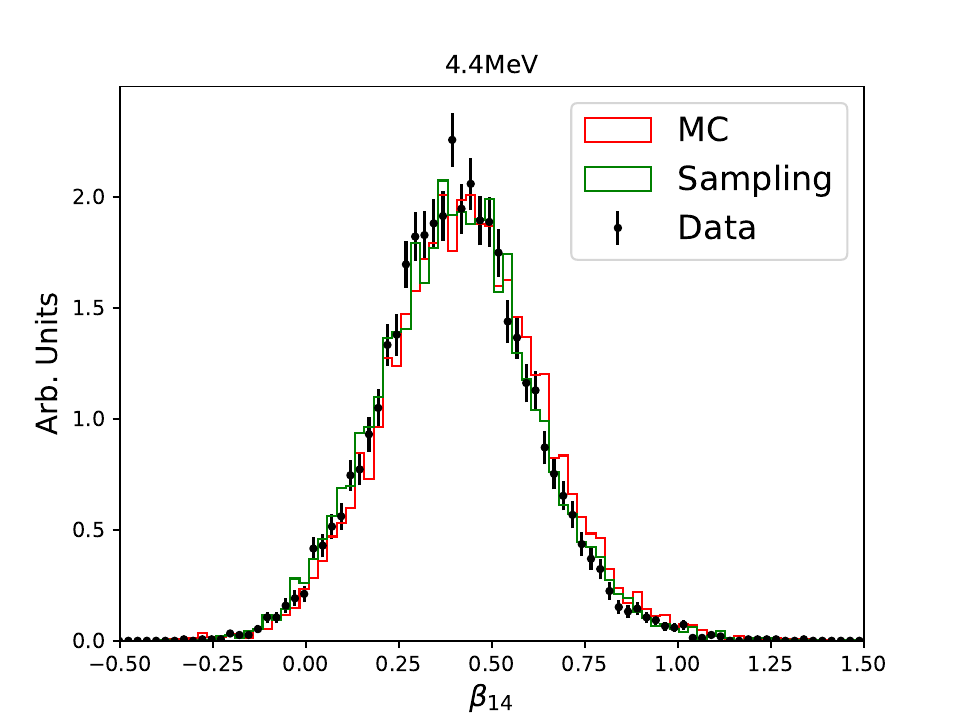}}
    \subfloat{\includegraphics[width=0.46\textwidth]{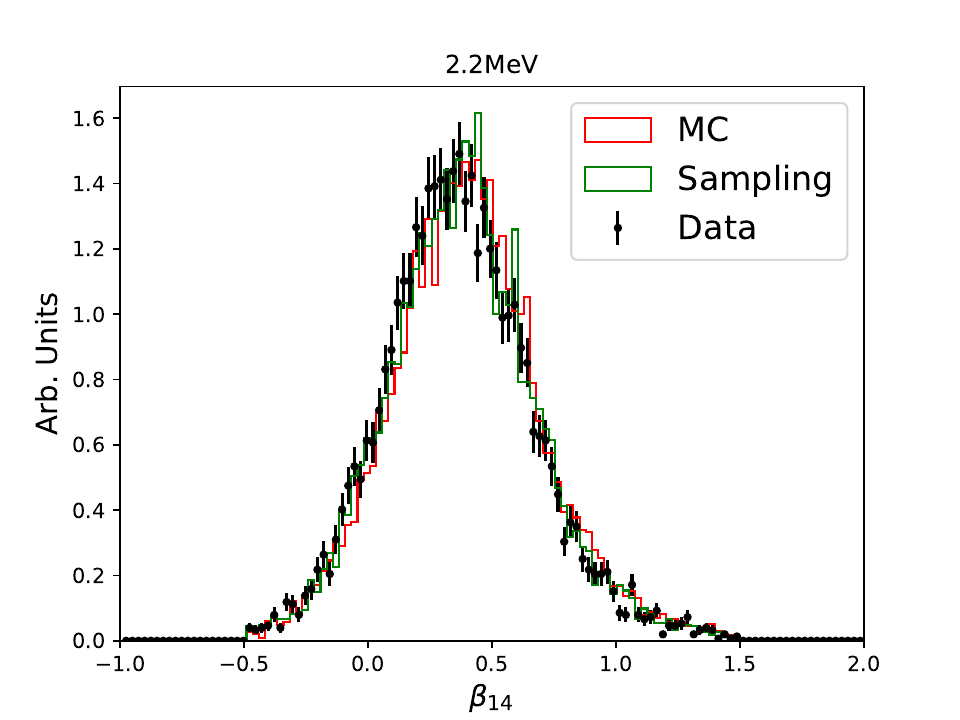}}
    \caption{Comparison of the $\beta_{14}$ distribution of the AmBe source, 2.2~MeV (left) and 4.4~MeV (right), for data, the default SNO+ MC and simulation produced using the Cherenkov light sampling method. }
    \label{fig:beta14_AmBe}
\end{figure*}

\begin{figure*}[!b]
    \centering
    \subfloat{\includegraphics[width=0.46\textwidth]{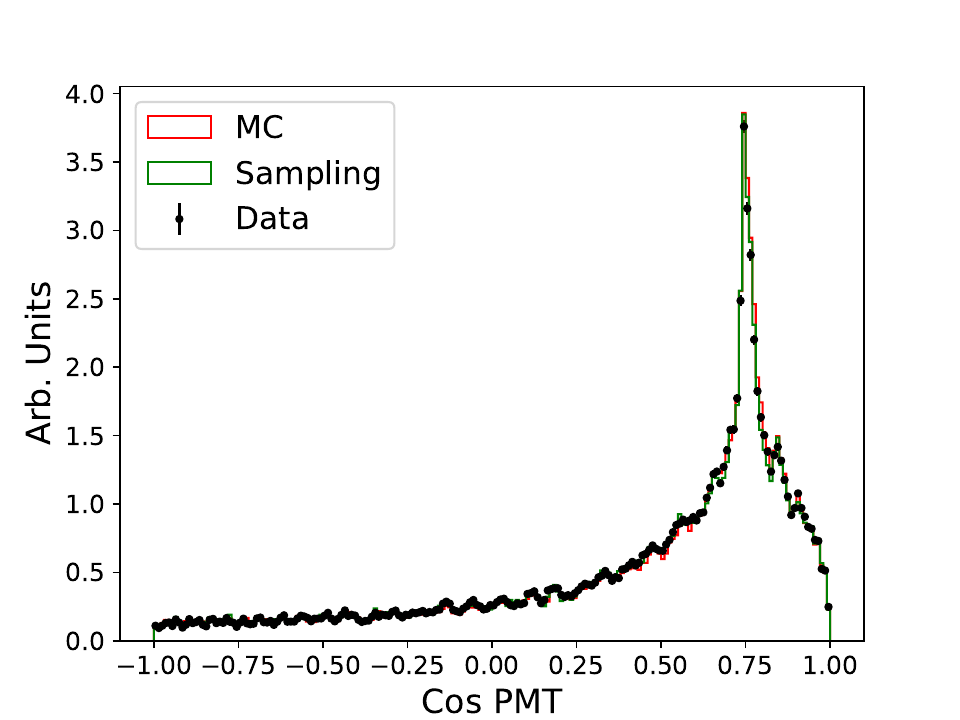}}
    \subfloat{\includegraphics[width=0.46\textwidth]{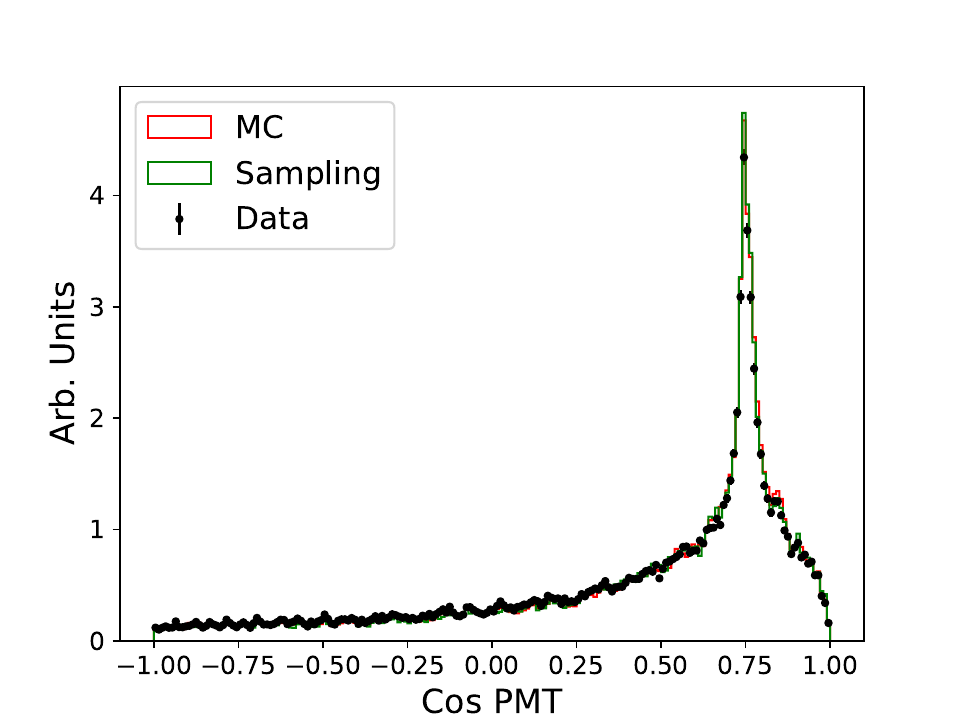}}
    \caption{Comparison of the $\cos\theta_{PMT}$ distribution of the AmBe source, 2.2~MeV (left) and 4.4~MeV (right), for data, the default SNO+ MC and simulation produced using the Cherenkov light sampling method. }
    \label{fig:CosPMT_AmBe}
\end{figure*}

The comparison of the tuned Cherenkov light sampling distributions and the SNO+ default simulation for AmBe data are shown below. The $\beta_{14}$ distribution is contained in Fig.~\ref{fig:beta14_AmBe}, while $\cos\theta_{PMT}$ can be found in Fig.~\ref{fig:CosPMT_AmBe}. We note that, after completing the analysis, the SNO+ collaboration found issues with the accuracy of the reconstructed event positions in the AmBe runs. Nonetheless, despite these issues, Figs.~\ref{fig:beta14_AmBe} and~\ref{fig:CosPMT_AmBe} display a better agreement with the Cherenkov sampling model than with the standard SNO+ simulation. Moreover, the $^{16}$N data dominates the final fit of the $\alpha$ parameter, and therefore we do not expect these issues to have an impact on our final result.

\end{document}